\documentclass[preprint,prb,showpacs,preprintnumbers,amsmath,amssymb,superscriptaddress]{revtex4}
\usepackage{graphicx}% Include figure files
\usepackage{dcolumn}% Align table columns on decimal point
\usepackage{bm}% bold math

\begin{document}

\title{High-performance $n$-type organic field-effect transistors with ionic liquid gates}

\author{S. Ono}

\altaffiliation{Electronic mail: shimpei@criepi.denken.or.jp}
\affiliation{DPMC and GAP, University of Geneva, 24 quai Ernest-Ansermet CH1211 Geneva,
Switzerland}
\affiliation{Central Research Institute of Electric Power
Industry, Komae, Tokyo 201-8511, Japan}
\author{N. Minder}
\affiliation{DPMC and GAP, University of Geneva, 24 quai Ernest-Ansermet CH1211 Geneva,
Switzerland}
\author{Z. Chen}
\affiliation{Polyera Corporation, 8025 Lamon Avenue, Skokie, Illinois 60077}
\author{A. Facchetti}
\affiliation{Polyera Corporation, 8025 Lamon Avenue, Skokie, Illinois 60077}
\author{A. F. Morpurgo}
\altaffiliation{Electronic mail: Alberto.Morpurgo@unige.ch}
\affiliation{DPMC and GAP, University of Geneva, 24 quai Ernest-Ansermet CH1211 Geneva,
Switzerland}

\date{\today}

\begin{abstract}
High-performance $n$-type organic field-effect transistors were developed with ionic-liquid gates and
N,N$^"$-bis(n-alkyl)-(1,7 and 1,6)-dicyanoperylene-3,4:9,10-bis(dicarboximide)s single-crystals. Transport measurements show that these devices reproducibly operate in ambient atmosphere with negligible gate threshold voltage and mobility values as high as  5.0 cm$^2$/Vs. These mobility values are essentially identical to those measured in the same devices without the ionic liquid, using vacuum or air as the gate dielectric. Our results indicate that the ionic-liquid and $n$-type organic semiconductor interfaces are suitable to realize high-quality $n$-type organic transistors operating at small gate voltage, without sacrificing electron mobility.
\end{abstract}

\maketitle
Organic field-effect transistors (OFETs) have attracted considerable attention because of their potential for applications in large-area, flexible, low-cost electronics.\cite{Malliaras} One of the major challenges for organic electronics is the realization of complementary circuits, where the simultaneous use of high-performance $n$- and $p$-type OFETs is essential. However, compared to $p$-type OFETs, $n$-type OFETs generally suffer from low carrier mobility and large positive threshold voltage, both of which pose serious problems for their practical use. The far-from-ideal performance of several $n$-type OFETs is due, in large part, to atmospheric oxidants such as O$_2$ and H$_2$O, and to chemical groups present at the organic-gate dielectric interface (e.g., hydroxyl groups in the case of the commonly used SiO$_2$ dielectrics) which act as electron traps thereby degrading transport properties.\cite{Friend} Because interfacial phenomena crucially determine the device performance,\cite{Veres,high-k} several combinations of gate dielectrics and organic semiconductors are being intensively tested\cite{Kitamura, Klauk, Walser} with the goal of optimizing $n$- and $p$-type conduction, as it is needed to achieve higher-performance complementary organic circuits.

Besides conventional solid-state dielectrics, an alternative method to accumulate carriers at the surface of organic semiconductors exploits the use of electric double layers (EDLs) at the interface with electrolytes. The advantage of electrolyte gating using EDL is that much higher carrier concentrations are accumulated by applying only low gate-bias voltages, because EDLs have a very large electrostatic capacitance owing to their small physical thickness (typically $\sim$ 1 nm). Indeed, for $p$-type transistors, the successful operation of high-quality OFETs based on EDL gating has been recently reported by several groups.\cite{Panzer1, Panzer2, Takeya, Shimotani, Lee, Misra, Cho} Among all investigated EDL-OFETs, room temperature ionic liquids (ILs) are particularly promising,\cite{Ono, Uemura, Ono2, Uemura2} as they have shown to enable a very fast response (compared to other types of electrolytes) to changes in the gate voltage. For $p$-type transistors (based on rubrene) it was also recently found that the surface of organic semiconductors is not damaged by the ILs, and mobility values up to 9.5 cm$^2$/Vs can be achieved by selecting the proper IL.\cite{Ono2} For $n$-type OFETs with EDL gating, however, virtually nothing is known at this stage.

In this paper our goal is to improve the performance of $n$-type OFETs and, by utilizing an EDL gating with ILs, we show that it is possible to reproducibly obtain high-quality devices exhibiting high mobility and ideal characteristics in air and operating at few volts.  To this end, we have fabricated N,N$^"$-bis(n-alkyl)-(1,7 and 1,6)-dicyanoperylene-3,4:9,10-bis(dicarboximide)s (PDIF-CN$_2$) single-crystal transistors with IL, using the same methods used for $p$-type rubrene single crystals.\cite{Uemura, Ono2} As organic material we have selected PDIF-CN$_2$,\cite{Facchetti, Facchetti2} which exhibits the highest electron mobility reported to date(1-6 cm$^2$/Vs).\cite{Molinari} We show that the electron mobility of PDIF-CN$_2$ single-crystal transistors with IL (as high as 5.0 cm$^2$/Vs in air) is comparable to the corresponding devices without IL, thus having air as gate dielectric. Our results demonstrate that even for electron-conducting materials, the use of ILs does not decrease the properties of the organic semiconductor, and that they are beneficial to improve several aspects of the device characteristics. We conclude that IL gating is a very promising approach enabling high-performance organic complementary circuits.

Figure 1 shows the device structure employed in this study (based on an elastomer stamps with a recessed gate -see Ref. [22] for details), as well as the IL used (N-methyl-N-propyl pyrrolidinium bis(trifluoromethanesulfonyl) imide ([P13][TFSI])), and its frequency-dependent capacitance. For rubrene OFETs it has been demonstrated that the charge carrier mobility is larger when the capacitance of the IL is smaller.\cite{Ono2} We have therefore selected [P13][TFSI] for our investigations, since this compound has the smallest capacitance among all ILs which we have worked with. Note that this compound is also expected to absorb less water as compared to others ILs, because of the strong hydrophobicity of the fluoroterminated anion, which makes it particularly favorable for investigating $n-$type OFETs. The frequency-dependent capacitance of the EDL ($C_{EDL}$) of [P13][TFSI] was measured on a test device consisting of polydimethylsiloxane (PDMS) substrates, similar to those used for the realization of the actual transistors, using Solartron 1260 and 1296 impedance analyzers (from 0.1 Hz to 10 MHz with a 5 mV AC voltage). Notably (see Fig. 1(c)), the value of the capacitance does not significantly decrease with increasing frequency, and it remains close to 1 $\mu$F/cm$^2$ even at 0.1 MHz. This implies that [P13][TFSI] enables the rapid accumulation of high-density carriers at the IL/semiconductor interfaces, which is particularly important for practical applications (i.e., the use of [P13][TFSI] does not limit the switching speed of the devices).

In our devices, the PDIF-CN$_2$ single crystals (grown by physical vapor transport) are laminated manually onto a PDMS stamp as shown in Fig. 1(a). The  gate, source, and drain electrodes consist of Ti/Au films; the distance between the crystal and the recessed gate is intentionally kept large (25 $\mu$m), to facilitate the insertion of the IL in the region between the gate and the crystal (when a small droplet is positioned near the device, the IL is "sucked in" spontaneously by capillary forces). An optical microscope image of one of our single crystal devices is shown in Fig. 1(b). An important advantage of this device structure is that we can measure the device characteristics before inserting the IL, which enables the direct comparison of the properties of the same transistor with air as gate dielectric, and with the EDL gate.

All the measurements were performed in air at room temperature, with an Agilent Technology E5270B semiconductor parameter analyzer. Before introducing the IL, the devices are characterized by measuring transfer and output characteristics (insets of Fig. 2(a),(c)). The PDIF-CN$_2$ single crystal devices show rather small --recall that the distance between gate and crystal is 25 $\mu$m -- threshold voltages of $\sim$ 20 V, unlike most other air-stable $n$-type organic transistors. The transfer characteristics show small hysteresis (inset Fig. 2(a)) whereas the hysteresis of the output characteristics is larger (inset of Fig. 2(c)). We measured approximately 15 devices and found that the mobility values $\mu_{air}$ with air as gate dielectric are in the range of 1.5 - 5.5 cm$^2$/Vs (these values are extracted from measurements performed in a four-terminal configuration to eliminate possible contact effects). These values are comparable to the best electron mobilities reported for PDIF-CN$_2$ single-crystal transistors.\cite{Molinari}

Figure 2(a) shows the transfer characteristics of PDIF-CN$_2$ single crystal OFETs after the IL [P13][TFSI] has been inserted between the gate and the crystal. The simultaneously measured gate leakage current $I_G$ through the electrolytes is
negligibly small as compared to the drain current $I_D$, that is less than 0.15 nA as long as $|V_G|$ is less than 0.5 V (see Fig. 2(b)).
It is apparent that the threshold voltage is drastically reduced to only 0.1 V. Gate voltages as low as 0.5 V are sufficient to obtain a large channel conductance. This observation directly demonstrates the ability to operate PDIF-CN$_2$ transistors with IL gates at very low voltages. From the transfer characteristics we estimate the normalized sheet transconductance $\sigma^T$ in the linear region, which directly determines the current amplification in OFETs.\cite{Uemura2} The obtained $\sigma^T$ reaches values as high as 6.1 $\mu S$ at $V_D$= 0.1 V, two orders of magnitude larger than the best $n$-type OFETs so far \cite{Uemura2} and comparable to the best value of $p$-type OFETs with IL gates.\cite{Uemura} The output characteristics of the PDIF-CN$_2$ single crystal OFETs with [P13][TFSI] are shown in Fig. 2(c). With increasing drain voltage $V_D$, the drain current $I_D$ first shows a linear increase, followed by a fairly good saturation behavior by applying less than 0.5 V for both $V_G$ and $V_D$. Noticeably, negligible hysteresis is observed PDIF-CN$_2$ single crystal OFETs with [P13][TFSI], even though in air-gap devices hysteresis in the same measurements is present. This indicates that ILs function does not only induce high-density carriers at the surface of the crystal, but also helps to eliminate aspects of device non-ideality.

Figure 3 shows the typical transfer characteristics of PDIF-CN$_2$ single crystal OFETs with [P13][TFSI] gates obtained in a four-terminal configuration. The conductivity measured for different values of $V_D$ lay on top of each other, indicating that the channel conductance is independent of source-drain voltage, as expected. The electron mobility $\mu_{IL}$ in IL devices (see Fig. 3 inset) is calculated from the conductivity measured in a four-terminal configuration, by employing the standard formula valid in the linear regime, and using the measured capacitance value (see Fig. 1(c)). The mobility increases with increasing $V_G$ and saturates above $V_G$ = 0.4 V, which we adopt as the value of $\mu_{IL}$. In different devices, the values of $\mu_{IL}$ are in the range of 1.0 - 5.0 cm$^2$/Vs, which are the highest ever reported for EDL-gated $n$-type OFETs. Figure 4 shows a comparison of $\mu_{IL}$ with $\mu_{air}$. Remarkably in 9 out of 15 devices the same mobility is observed with IL gates, and when using air as a gate dielectric (in the remaining samples the mobility with ionic liquid is slightly smaller than, but comparable to, that measured with air as gate dielectric). Importantly, this observation indicates that the surface of the PDIF-CN$_2$ single crystals is not affected by the contact with the IL: it follows that by choosing the proper ILs, we can benefit from the intrinsic semiconductor properties while operating the devices at much lower voltages.

In conclusion, we have fabricated $n$-type single-crystals OFETs with ionic liquid gates. By choosing the most suitable ionic liquid, we have shown that these devices can combine top performance in terms of mobility values (up to 5 cm$^2$/Vs), threshold voltage ($<$ 0.1 V), and ideality of the electrical characteristics, with the possibility of very low-voltage operation. These results are particularly relevant for implementation of $n$-type OFETs into complementary logic architectures for realizing integrated circuits based on organic transistors.

The authors acknowledge J. Takeya and C. D. Frisbie for helpful discussion and
K. Miwa, M. Nakano and I. G. Lezama for their technical assistance.
The study was supported  by FCT, Grant-in-Aid for Scientific Research (Grant No. 20740213) from MEXT, NEDO and MaNEP.

\newpage

\begin{figure}
\includegraphics[clip,width=7.0cm]{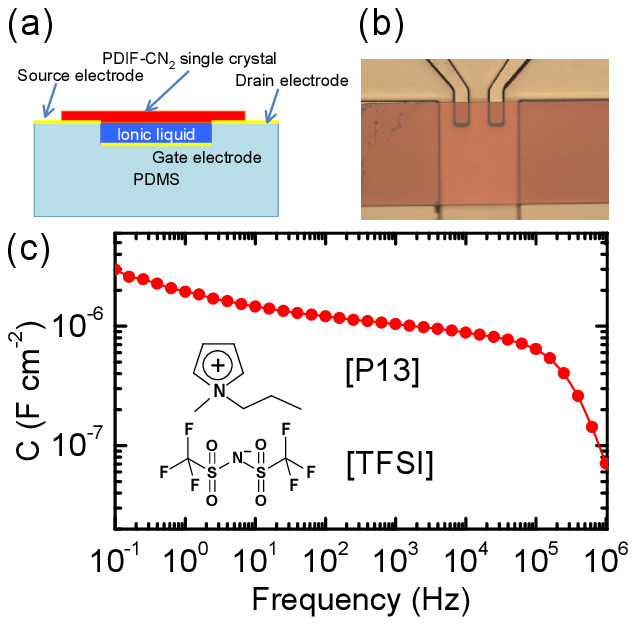}
\caption{(color online)
(a) Schematic structure of the ionic liquid gated devices. (b) Optical microscope image of a PDIF-CN$_2$ crystal/[P13][TFSI] transistor. (c) Frequency dependence of the capacitance of [P13][TFSI] EDL, measured by the ac impedance technique. (Inset) Chemical structures of [P13][TFSI].}
\end{figure}

\begin{figure}
\includegraphics[clip,width=7.0cm]{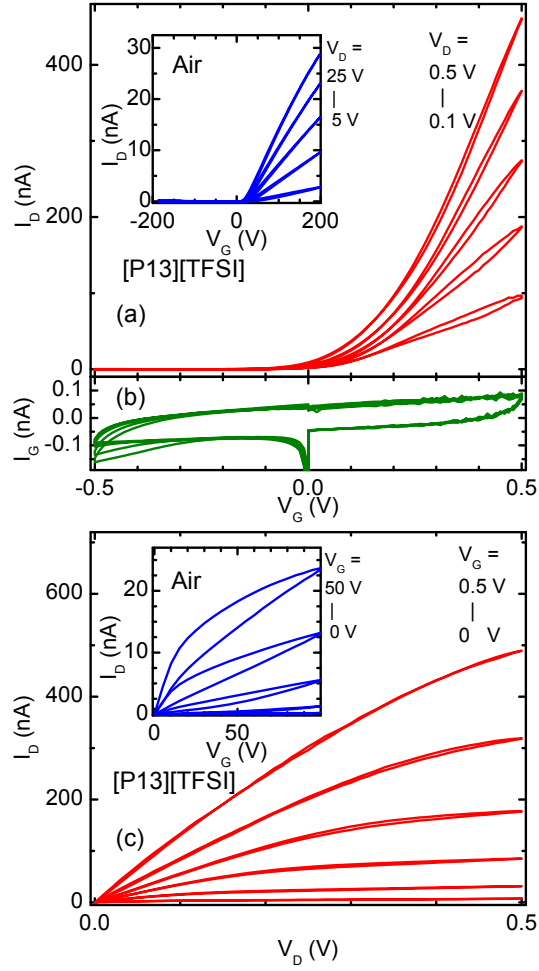}
\caption{(color online)
(a) Transfer characteristics and (b) gate leakage current $I_G$ of a PDIF-CN$_2$/[P13][TFSI] single-crystal OFET measured with $V_D$ = 0.1, 0.2, 0.3, 0.4, 0.5 V. (c) Output characteristics of the same device measured with $V_G$ = 0, 0.1, 0.2, 0.3, 0.4, 0.5 V. The insets in (a) and (c) show measurements on the same device before inserting the IL (in (a):  $V_D$ = 5, 10, 15, 20, 25 V; in  (c) $V_G$ = 0, 10, 20, 30, 40, 50 V).}
\end{figure}

\begin{figure}
\includegraphics[clip,width=7.0cm]{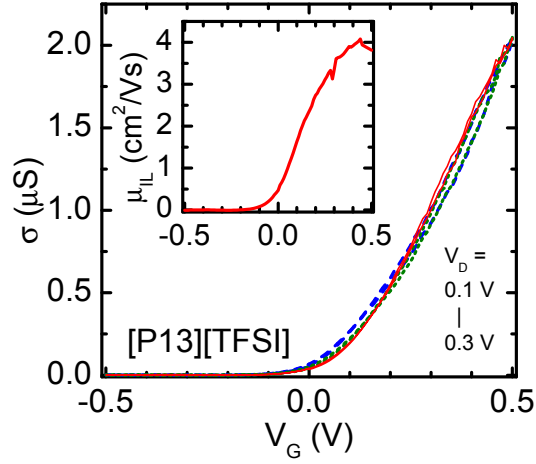}
\caption{(color online)
Gate-voltage dependence conductivity of a [P13][TFSI]/PDIF-CN$_2$ single crystal OFETs measured in a four-terminal configuration with different drain voltages ($V_D$ = 0.1 (red), 0.2 (Green), 0.3 V (Blue)). The corresponding carrier mobility as a function of gate voltage is shown in inset.}
\end{figure}

\begin{figure}
\includegraphics[clip,width=7.0cm]{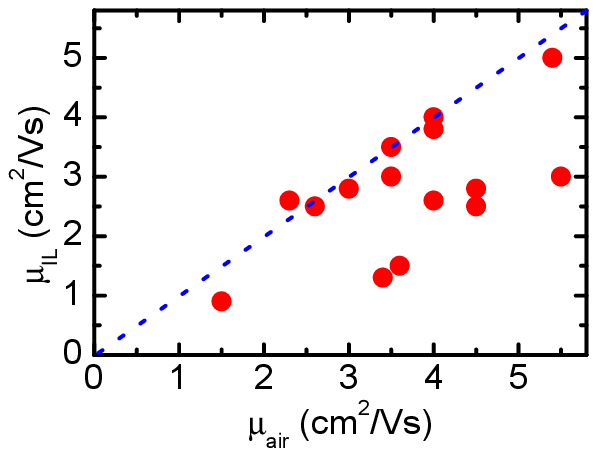}
\caption{(color online)
Comparison of the mobility of [P13][TFSI]/PDIF-CN$_2$ single crystal OFETs with that obtained from the same devices before inserting the IL. }
\end{figure}


\begin{references}
\newpage

\bibitem{Malliaras}
G. Malliaras and R. Friend,
Phys. Today {\bf 58}, 53 (2005).

\bibitem{Friend}
L.-L. Chua, J. Zaumseil, J.-F. Chang, E. C.-W. Ou, P. K.-H. Ho, H. Sirringhaus, and R. Friend,
Nature {\bf 434}, 194 (2005).

\bibitem{Veres}
J. Veres, S. D. Ogier, S. W. Leeming, D. C. Cupertino, and S. M.  Khaffaf,
Adv. Funct. Mater. {\bf 13}, 199 (2003).

\bibitem{high-k}
A.F. Stassen, R.W.I. de Boer, N.N. Iosad, and A.F. Morpurgo, Appl. Phys. Lett. {\bf 85}, 3899 (2004);
I.N. Hulea, S. Fratini, H. Xie, C.L. Mulder, N.N. Iossad, G. Rastelli, S. Ciuchi, and A.F. Morpurgo,
Nat. Mat. {\bf 5}, 982 (2006).

\bibitem{Kitamura}
M. Kitamura and Y. Arakawa,
Appl. Phys. Lett. {\bf 91}, 053505 (2007).

\bibitem{Klauk}
H. Klauk, U. Zschieschang, J. Pflaum, and M. Halik,
Nature {\bf 445}, 745 (2007).

\bibitem{Walser}
M. P. Walser, W. L. Kalb, T. Mathis, T. J. Brenner, and B. Batlogg,
Appl. Phys. Lett. {\bf 94}, 053303 (2009).

\bibitem{Panzer1}
M. J. Panzer and C. D. Frisbie,
J. Am. Chem. Soc. {\bf 127}, 6960 (2005).

\bibitem{Panzer2}
M. J. Panzer and C. D. Frisbie,
Appl. Phys. Lett. {\bf 88}, 203504 (2006).

\bibitem{Takeya}
J. Takeya, K. Yamada, K. Hara, K. Shigeto, K. Tsukagoshi, S. Ikehata, and Y. Aoyagi,
Appl. Phys. Lett. {\bf 88}, 112102 (2006).

\bibitem{Shimotani}
H. Shimotani, H. Asanuma, J. Takeya, and Y. Iwasa,
Appl. Phys. Lett. {\bf 89}, 203501 (2006).

\bibitem{Lee}
J. Lee, M. J. Panzer, Y. He, T. P. Lodge, and C. D. Frisbie,
J. Am. Chem. Soc. {\bf 129}, 4532 (2007).

\bibitem{Misra}
R. Misra, M. McCarthy, and A. F. Hebard,
Appl. Phys. Lett. {\bf 90}, 052905 (2007).

\bibitem{Cho}
J. H. Cho, J. Lee, Y. Xia, B. Kim, Y. He, M. J. Renn, T. P. Lodge, and C. D. Frisbie,
Nature Mat. {\bf 7} 900 (2008).

\bibitem{Ono}
S. Ono, S. Seki, R. Hirahara, Y. Tominari, and J. Takeya,
Appl. Phys. Lett. {\bf 92}, 103313 (2008).

\bibitem{Uemura}
T. Uemura, R. Hirahara, Y. Tominari, S. Ono, S. Seki, and J. Takeya,
Appl. Phys. Lett. {\bf 93}, 263305 (2008).

\bibitem{Ono2}
S. Ono, K. Miwa, S. Seki, and J. Takeya,
Appl. Phys. Lett. {\bf 94}, 063301 (2009).

\bibitem{Uemura2}
T. Uemura, M. Yamagishi, S. Ono, and J. Takeya,
Appl. Phys. Lett. {\bf 95}, 103301 (2009).

\bibitem{Facchetti}
B. A. Jones, M. J. Ahrens, M. -H. Yoon, A. Facchetti, T. J. Marks, and M. R. Wasielewski,
Angew. Che,. Int. Ed. {\bf 43}, 6363 (2004).

\bibitem{Facchetti2}
T. Jung, B. Yoo, L. Wang, A. Dodabalapur, B. A. Jones, A. Facchetti, M. R. Wasielewski, and T. J. Marks,
Appl. Phys. Lett. {\bf 88}, 183102 (2006).

\bibitem{Molinari}
A. S. Molinari, H. Alves, Z. Chen, A. Facchetti, and A. F. Morpurgo,
J. Am. Chem. Soc. {\bf 131}, 2462 (2009).

\bibitem{Menard}
E. Menard, V. Podzorov, S.H. Hur, A. Gaur, M.E. Gershenson, and J.A. Rogers, Adv. Mat. {\bf 16}, 2097 (2004).

\end{references}
\end{document}